\documentclass{PoS}
\usepackage[dvips]{epsfig}
\usepackage{subfigure}

\title{Taste symmetry violation at finite temperature }

\ShortTitle{Taste symmetry violation at finite temperature }

\author{\speaker{Fabrizio Pucci}\\
        Fakult\"{a}t f\"{u}r Physik, Universit\"{a}t Bielefeld, D-33615 Bielefeld, Germany\\
        E-mail: \email{pucci@physik.uni-bielefeld.de}}

\author{Edwin Laermann\\
        Fakult\"{a}t f\"{u}r Physik, Universit\"{a}t Bielefeld, D-33615 Bielefeld, Germany\\
        E-mail: \email{edwin@physik.uni-bielefeld.de }}

\abstract{Symmetries play a distinctive role at the high temperature phase transition in QCD. Therefore the spectrum of screening masses has been investigated with emphasis on taste breaking. Although taste violation is an UV effect the relevant operators could be temperature dependent. We have studied the meson screening masses in the temperature range between 140 MeV to 550 MeV. The computation has been performed using dynamical $N_f$ = 2+1 gauge field configurations generated with the p4 staggered action. For temperatures below the transition an agreement with the prediction of staggered chiral perturbation theory has been found and no temperature effect can be observed on the taste violation. Above the transition the taste splitting still shows an $O(a^2)$ behavior but with a temperature dependent slope.}

\FullConference{The 30th International Symposium on Lattice Field Theory\\
                 June 24 - 29,  2012\\
                 Cairns, Australia}

\begin{document}

\section{Introduction}          \label{se.intro}
\noindent
One of the most used formulation of
lattice quantum chromodynamics (LQCD) at finite temperature is the staggered fermion formulation introduced
for the first time in \cite{Kogut:1974ag} by Kogut and Susskind. Despite the fact that it can not solve completely the fermion doubling problem it is computationally cheaper and preserves a remnant of the chiral symmetry.\newline
For every quark introduced in the theory one has four states that are usually called tastes
to distinguish them from the ordinary flavors. In the continuum all the tastes are degenerate but they split at
non-zero lattice spacing. In order to obtain the correct number of degrees of freedom a rooting procedure has to be
introduced.\newline
Just by counting, in a theory with $N_f$ number of flavors, there are $16 N_f^2$-1 pion states but only $N_f^2-1$ pions are the true
Goldstone bosons while the remaining states are unphysical. The presence of these states, that are classified
according to their transformation properties under the symmetry group preserved by the
staggered formulation \cite{Golterman:1985dz,Aubin:2004fs,Bazavov:2010pg}, contributes to the cut-off dependence of physical
quantities computed on the lattice. In order to have a better taste symmetry there are two possibilities, on one side
one can decrease the lattice spacing and go to larger temporal extent in
finite-temperature calculations or one can use an action with higher degree of improvement \cite{Heller:1999xz,Karsch:2000ps,Orginos:1999cr,Aoki:2005vt,Follana:2006rc}.
\newline
In order to understand these cut-off effects due to taste symmetry breaking one usually looks at the so called taste splitting
defined as

\begin{equation}\Delta_{\xi} = m_{[\xi]}^2 - m_{\xi_5}^2,\label{taste}\end{equation}
\noindent
where $m_{\xi_5}$ is the mass of the Goldstone boson and $m_{[\xi]}$ is the mass of a taste state labeled by the taste matrix $\xi$.
At zero temperature the lattice results (see \cite{Aubin:2004fs,Bazavov:2010pg,Bazavov:2010ru} for the most recent investigations)
indicate that the taste splittings in the pion sector go like $a^2$ and are in agreement with staggered chiral perturbation theory
\cite{LeeSharpe,AubinBernard,Bernard:2007qf}.
\newline
At finite temperature where S$\chi$PT is in principle not reliable the situation could be different. In the confined phase
one could expect a modification of the chiral Lagrangian leading for example to
temperature dependent low energy effective constant (LECs). Moreover at temperatures above the transition
S$\chi$PT is not valid and the situation has to be better understood.
\newline
Here we address these problems studying the dependence of taste breaking on lattice spacing
as well as temperature. Using the dynamical $N_f$=2+1 gauge field configurations
generated with the RHMC algorithm by the RBC-Bielefeld \cite{RBCBi-eos}
and the HotQCD \cite{hotQCD} collaborations using the p4 staggered action we will measure the taste splitting (\ref{taste}) in a range of temperature from about $140$ MeV to $550$ MeV focusing on the pion sector. In order to analyze the lattice spacing dependence of these quantities two different sets of the lattice spacing corresponding to $N_\tau= 6$ and $8$ will be used.\newline
In the following, after recalling some basic knowledge of staggered formulation, we will present the results regarding the temperature dependence of taste splitting at fixed lattice spacing and the lattice spacing dependence of that quantity having instead fixed $T$. A more general analysis
and further information regarding the taste breaking in other mesonic channels has been presented in \cite{Laermann:2012sr}.

\section{Staggered Mesons}
\noindent
The staggered mesons are built from the tensor product of two staggered fields and can be divided in two different classes. The first are the local operators where the quark $\chi(x)$ and the anti-quark $\bar{\chi}(y)$ sit at the same lattice point. They can be written as

\begin{equation}M_{local} = \phi(\textbf{x}) \bar{\chi}(\textbf{x})\chi(\textbf{x}),\end{equation}
\noindent
with $\phi(\textbf{x}$) being a phase factor depending on the choice of the mesonic channel. The other operators, namely the one, two, and three-link operators, defined respectively as

\begin{equation}M_{one-link} = \phi(\textbf{x}) \bar{\chi}(\textbf{x})\Delta_i \chi(\textbf{x})\end{equation}
\begin{equation}M_{two-link} =\epsilon_{ijk} \phi(\textbf{x}) \bar{\chi}(\textbf{x})\Delta_i \Delta_j\chi(\textbf{x})\end{equation}
\begin{equation}M_{three-link} = \phi(\textbf{x}) \bar{\chi}(\textbf{x})\Delta_1 \Delta_2 \Delta_3 \chi(\textbf{x})\end{equation}
\noindent
are non-local since the two staggered fields don't sit at the same point but are shifted by the operator

\begin{equation}\Delta_i\, \chi(\textbf{x}) = 1/2 ( \chi(\textbf{x}+ \hat{i}) +  \chi(\textbf{x}- \hat{i})).\end{equation}
\noindent
From all these states we can select and study the sixteen pion operators that consist of two local operators, six one-link, six two-link and two three-link operators (for more details see \cite{Golterman:1985dz,altmeyer}). Every mesonic correlator
contains two different states with opposite parity. It can be parametrized as
\begin{eqnarray}
C(z) &=& A_{NO} \cosh\left[ M_- \left( z - \frac{N_s}{2} \right)\right] - (-1)^z A_{O} \cosh\left[ M_+ \left( z - \frac{N_s}{2} \right)\right] .
\label{eq.cor-par}
\end{eqnarray}
\noindent
where $A_{NO}$ and $A_O$, $M_+$ and $M_-$ are the amplitude and the masses of the two states. The pion states will thus appear as the oscillating or the non-oscillating contributions of such correlators. From general group theory arguments one can show that these sixteen states are degenerate in the continuum limit and split at non-zero lattice spacing. The pattern of this splitting can be read as

\begin{equation}16 \rightarrow \underbrace{1 \oplus 1}_{local} \oplus \underbrace{3 \oplus 3}_{one-link} \oplus \underbrace{3 \oplus 3}_{two-link} \oplus \underbrace{1 \oplus 1}_{three-link}.\end{equation}
\noindent
Having data for all the different states it is possible to define the so-called root-mean-squared (RMS) pion mass\footnote{In the spinor-taste basis we will
indicate with $\gamma$ and $\xi$ the spin and the taste gamma matrices respectively.} and study it as a function of the lattice spacing

\begin{equation}m_{\pi}^{RMS} = \sqrt{\frac{1}{16} \left( m_{\xi_5}^2 + m_{\xi_0 \xi_5}^2  + 3 m_{\xi_i \xi_5}^2  + 3 m_{\xi_i \xi_j}^2  + 3 m_{\xi_i \xi_0}^2 + 3 m_{\xi_i}^2  + m_{\xi_0}^2  + m_{I}^2 \right)}.\end{equation}
\noindent
However from the lattice data at zero temperature (see for example \cite{Aubin:2004fs,Bazavov:2010pg,Bazavov:2010ru}), one can see that the pion falls into only five classes following this pattern

\begin{equation}16 \rightarrow 1 \oplus 4 \oplus 4 \oplus 6 \oplus 1\end{equation}
\noindent
instead of eight as predicted by group theory.\newline This fact can be well explained with Staggered Chiral Perturbation Theory ( S$\chi$PT). Indeed if we focus only on the non-diagonal flavor pions\footnote{Simulations in which disconnected contributions are not taken into account describe only non-diagonal flavor states $\pi^{\pm}, K^{\pm}..$}, one can expand the staggered chiral lagrangian and find the tree-level masses of the pions that are given by
\noindent
\begin{equation}
m_{M_B}^2 = \mu ( m_a + m_b ) + a^2 \Delta_{\xi_B}
\label{eq.splittings}
\end{equation}
\noindent
where the meson $M$ is composed of two quarks $a$ and $b$ and where

\begin{equation}\Delta(\xi_5) = \Delta_{PS} = 0\label{a1}\end{equation}

\begin{equation}\Delta(\xi_{\mu 5}) = \Delta_{A} = \frac{16}{f_{\pi}^2} \left( C_1 + 3 C_3 + C_4 + 3 C_6 \right)\label{a2}\end{equation}

\begin{equation}\Delta(\xi_{\mu \nu}) = \Delta_{T} = \frac{16}{f_{\pi}^2} \left( 2 C_3 + 2 C_4 + 4 C_6 \right)\label{a3}\end{equation}

\begin{equation}\Delta(\xi_{\mu}) = \Delta_{V} = \frac{16}{f_{\pi}^2} \left( C_1 + C_3 + 3 C_4 + 3 C_6 \right)\label{a4}\end{equation}

\begin{equation}\Delta(\xi_I) = \Delta_{I} = \frac{16}{f_{\pi}^2} \left( 4 C_3 + 4 C_4 \right)\label{a5}\end{equation}
\noindent
so that \emph{e.g.} the $\xi_0 \xi_5$ and $\xi_i \xi_5$ taste states are degenerate. As we can understand from these splittings, in the pion sector there is a partial restoration of the taste symmetry. All the states fall into five degenerate multiplets according to the $SO(4)$ symmetry group. \newline
While this analysis is correct at zero temperature, it is not more valid at finite $T$ since other terms and temperature dependent LECs could appear in the chiral lagrangian.

\section{Result}
\begin{figure*}[!ht]
\begin{center}
\subfigure[]{\label{Pion32r0}\includegraphics[scale=0.8]{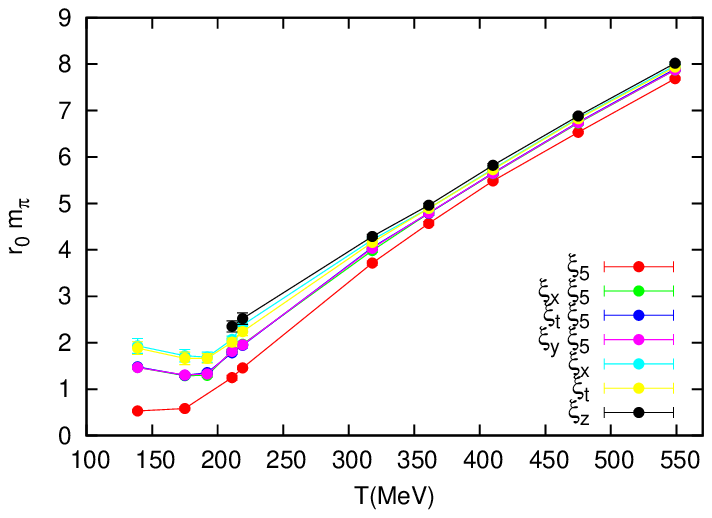}}
\hspace{1cm}
\subfigure[]{\label{Pion32T}\includegraphics[scale=0.8]{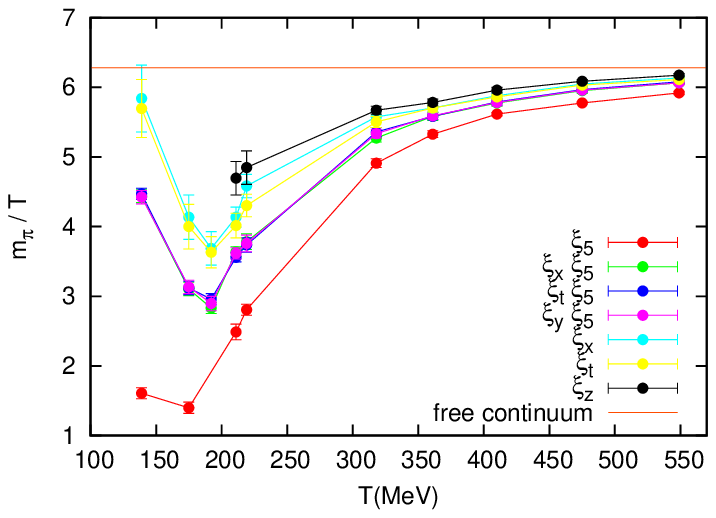}}
\subfigure[]{\label{pion24r0}\includegraphics[scale=0.8]{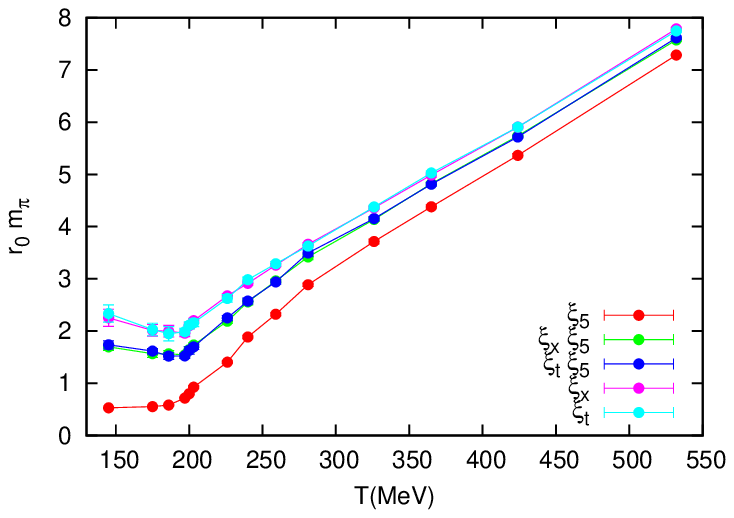}}
\hspace{1cm}
\subfigure[]{\label{pion24T}\includegraphics[scale=0.8]{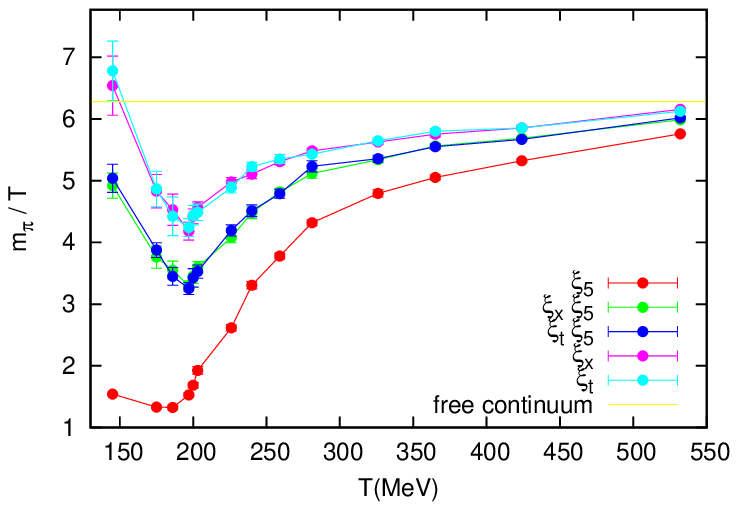}}
\end{center}
\caption{Temperature dependence of the pseudoscalar screening mass ($\gamma_5 \otimes [\xi]$)
in units of $r_0$ and of the temperature at $N_\tau = 8$ $(a,b)$ and $N_\tau = 6$ $(c,d)$ for different components of the multiplet identified by the taste matrix $[\xi]$.}
\label{fig.pion32}
\end{figure*}

\noindent
Here we will present the screening mass of all the pion states at finite $T$, both below and above the transition, trying to understand how the taste symmetry breaking occurs in the range 140 MeV $ < T <$ 550 MeV. Let us start with the pictures \ref{Pion32r0} and \ref{pion24r0} where the masses are plotted in units of $r_0$ as function of the temperature at $N_{\tau}=8$ and $N_{\tau}=6$ respectively.  While below the transition the Goldstone boson screening mass (the local operator) remains approximatively
constant, a slight decrease of all the other masses can be observed. Above the transition,
for all the channels a linear rise of the values of the masses occurs. This can be seen also in the figures \ref{Pion32T} and \ref{pion24T} where the screening masses are plotted in unit of $T$. At very high temperature their values approach from below the free continuum result given by $2 \pi T$.\newline
Unfortunately, differently from the zero temperature case in which all the taste states can be studied, at finite temperature there are some problems in the analysis due to the fact that the amplitude in some channels (where they appear in pairs with other non-pseudoscalar states) dies out very fast with rising temperature and as a consequence it is extremely difficult to extract their screening masses. As a consequence we can not
check a complete restoration of the $SO(4)$ taste symmetry as predicted by the S$\chi$PT for the zero temperature case since we don't have acces to all the states. However we have observed that at least an $SO(3)$ taste restoration occurs in the pion sector. Indeed, as we can see from the pictures, no difference between the $\xi_x$ and $\xi_t$ pions as well between $\xi_5 \xi_x$ and $\xi_5 \xi_t$ can be observed. From group theory arguments in principle the symmetry under which the multiplets should be classified at finite temperature is $SO(2) \times Z_2$.\newline

\begin{figure*}[!ht]
\begin{center}
\subfigure[]{\label{beta343}\includegraphics[scale=1.0]{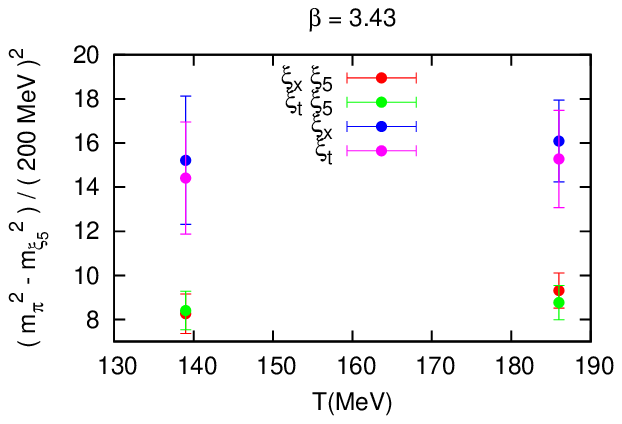}}
%\subfigure[]{\label{beta350}\includegraphics[scale=0.9]{betavst02.eps}}
\hspace{1cm}
\subfigure[]{\label{beta357}\includegraphics[scale=1.0]{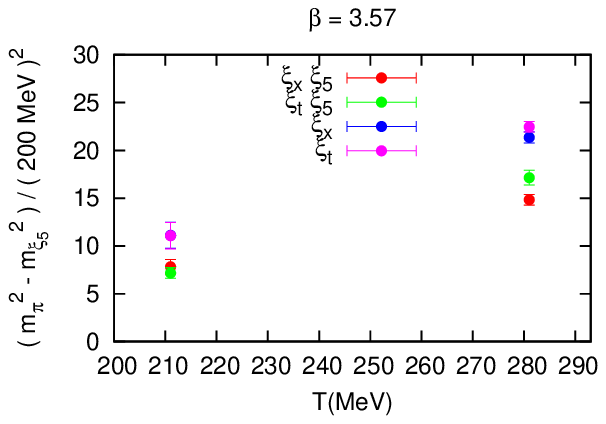}}
%\subfigure[]{\label{beta376}\includegraphics[scale=0.9]{betavst04.eps}}
%
%\subfigure[]{\label{beta392}\includegraphics[scale=0.9]{betavst05.eps}}
\end{center}
\caption{Effect of the temperature on the taste violation at fixed $\beta$ values : (a) $\beta = 3.43$, (b) $\beta = 3.57$}
\label{fig.betavst}
\end{figure*}

\begin{figure*}[!ht]
\begin{center}
\subfigure[]{\label{tvsafinalff}\includegraphics[scale=1.05]{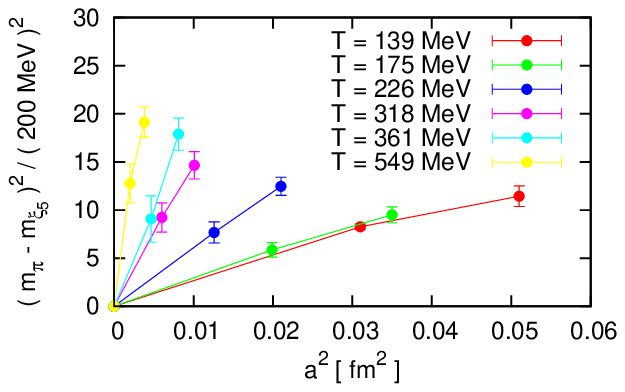}}
\hspace{0.4cm}
\subfigure[]{\label{tvsafinalf}\includegraphics[scale=1.05]{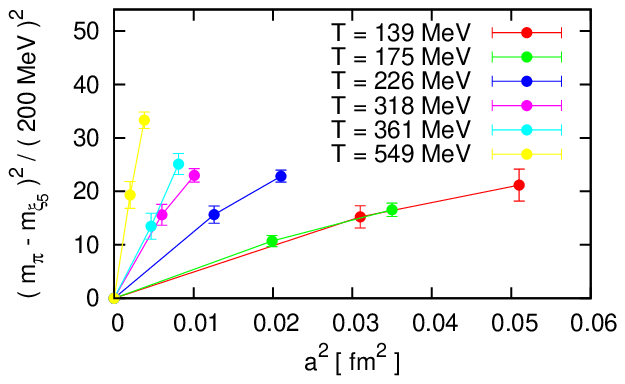}}
\end{center}
\caption{Lattice spacing dependence of the taste splitting for one link ($m_{\gamma_{5}\otimes \xi_x \xi_5}^2 - m_{\gamma_{5}\otimes \xi_5}^2$)(a) and two link meson operators ($m_{\gamma_{5}\otimes \xi_x }^2 - m_{\gamma_{5}\otimes \xi_5}^2$ )(b) at different temperatures. }
\label{fig.fff2}
\end{figure*}

\noindent
In figure (\ref{fig.betavst}), at fixed value of the lattice spacing the effect of the temperature on the taste splitting for the one-link and two link operators is reported while in picture (\ref{fig.fff2}) we fixed the temperature and consider how the taste splitting
depends on the lattice spacing. From both figures we can conclude that below the transition the taste splitting seems independent of the temperature. Above $T_c$ the situation changes and the taste splitting acquires a temperature dependence. The rising of the Goldstone pion masses with the temperature could explain this unexpected behavior of the taste splitting. Unfortunately it was not possible to establish the precise functional behavior of this dependence probably because we are in the transition region, where the $\xi_5$ pion is no longer a Goldstone boson and at the same time we are far from the free theory regime.

\section*{Acknowledgments}
\noindent
F.P. is supported by the Research Executive Agency (REA)
of the European Union under Grant Agreement PITNGA-
2009-238353 (ITN STRONGnet). The numerical
computations have been carried out on the apeNEXT at Bielefeld University.

\end{document}